\documentstyle[aps]{revtex}
\begin{document}
\title{Finite-temperature correlations in the trapped Bose gas}
\author{N.M. Bogoliubov${}^{\star }$, R.K. Bullough${}^{\dagger }$, V.S. Kapitonov$%
{}^{\star \star }$, C. Malyshev${}^{\star }$ and J. Timonen${}^{\dagger
\dagger }$}
\address{${}^{\star }$Steklov Institute of Mathematics at St.Petersburg, Fontanka
27, 191011 St.Petersburg, Russia}
\address{${}^{\dagger }$Department of Mathematics, The University of Manchester
Institute of Science and Technology, P.O. Box 88, Manchester M60 1QD,
United Kingdom}
\address{${}^{**}$St.Petersburg Technological Institute, Moskovsky 26, 196013, St.Petersburg, Russia}
\address{${}^{\dagger\dagger }$Department of Physics, University of Jyv\"askyl\"a,
P.O. Box 35, FIN-40351 Jyv\"askyl\"a, Finland}
\maketitle

\begin{abstract}
Previous functional integral methods for translationally invariant systems 
have been extended to the case of a confining trap potential. Essentially 
all finite-temperature properties of the repulsive Bose gas in a paraboloidal
trap can be determined this way. New analytical results reported here are for the 
finite-temperature two-point correlation functions below the critical 
temperatures in $d=3,2$ and $1$ dimensions. Only for $d=3$ are correlations 
both long range and coherent - at long range consistent with the existence
of a coherent state of the condensate in the trap. The two-dimensional
condensate is marginally stable in the sense that correlations decay by
a power law.

03.75.Fi, 05.30.Jp
\end{abstract}

The observation of Bose condensation in vapours of alkali atoms \cite{1,2,3}
held in magnetic traps, and recently \cite{4} in atomic hydrogen, has
stimulated enormous interest, both experimentally and theoretically, in this
phenomenon. The coherence properties of condensates in such traps are of
exceptional interest especially perhaps because of the opportunity now
presented for realisation of atom lasers \cite{5,6,Hansch}. Much of the theoretical
work has been concerned with solutions of the c-number Gross-Pitaevskii (GP)
equation in the presence of the paraboloidal potential describing the trap 
\cite{7,8,9}. Without the trap this equation is also called the Nonlinear
Schr\"odinger (NLS) equation \cite{10} which, in one space dimension ($d=1$%
), can be solved exactly at both c-number and quantum level, the latter
including \cite{11,bbt} calculation of the finite-temperature correlation
functions for the repulsive case (coupling constant $g>0$). For $d=1$ as
well as for the higher dimensions, the finite-temperature properties of the
quantum NLS equation have otherwise been extensively analysed \cite{pop1} by
functional integral methods. This way it is established e.g. that without
any trap no long-range correlations arise for $d<3$. But, on the other hand,
for $d=3$ correlations are both long-range and consistent with the
condensate being in a coherent state \cite{13}.

Thus far it has remained an open question whether the presence of a trap
potential will induce long-range order for $d<3$ also, while, despite \cite
{5,6}, coherent-state behaviour for $d=3$ in traps had not been demonstrated
theoretically until the very recent work of \cite{Glauber}. In \cite{Glauber}
first and second order coherence functions are calculated from the GP equation
in the form of series expansion. In this Letter we give an {\it ab initio}
approximate calculation of the first order coherence function for $d=3,2$, and
1, giving explicit analytical formulae in each case.
We show that, even in the presence of a
paraboloidal trap, there is long-range order only for $d=3$; and we indicate
how, and only for $d=3$, the condensate in the trap may, but only
asymptotically, be approaching a coherent state. To these ends we have extended
the previous \cite{pop1,pop2} functional integral methods to the case when a
trap is included. An important new aspect is then that the trap breaks
translational invariance; and this introduces wholly new many-body
theoretical problems. Thus, rather than using periodic boundary conditions
(b.c.s), we must expect to impose vanishing b.c.s at infinity, and so to
work at 'zero density' \cite{14}. In this Letter we simply divide \cite
{fet,fp} the total field $\psi $ of the system into two parts: $\psi =\psi
_0+\psi _1$. Here $\psi _0$ becomes for $T<T_c$ the condensate field and $%
\psi _1$ is that due to thermal fluctuations. Both fields are assumed to
vanish at infinity in order to have them largely confined in the region of
the trap. The functional integral method still provides a framework within
which, in principle, all thermodynamic properties of the trapped bose gas
can be determined. So far we have obtained, e.g., expressions for $T_c$ for
each of $d=3,2,1$. At mean-field level results at lowest order in the
coupling constant agree with those reported \cite{7} already, but they can
be extended to higher order. However, the main result reported in this
Letter is the calculation of the finite-temperature two-point correlation
functions for each of $d=3,2,1$. Because there
is no translational invariance these correlation functions no longer depend
solely on the difference of two position vectors. It can still be concluded
that long-range order arises, for $T<T_c$, in $d=3$, and the first-order
coherence function asymptotically approaches unity. In $d=2$ and for $T<T_c$
the condensate is {\it marginally} stable for correlations decay
algebraicly, namely as a power law. In $d=1$ correlations decay
exponentially for $T<T_c$, and we have not yet analysed any $T=0$ limit. For 
$T>T_c$ there is a Gaussian decay in all dimensions.

Our calculational procedure is such that we first draw a correspondence
between the quantum fields $\psi ({\bf r},\tau ),\psi ^{\dagger }({\bf r}%
,\tau )$ for Bose particles  and the c-number fields $\psi ({\bf r},\tau ),%
\bar \psi ({\bf r},\tau )$ ($\tau $ is the Wick rotated time $t\to -i\tau ,$
and varies between $0$ and $\beta =(k_BT)^{-1}$ ). In this way we can write
the finite-temperature correlation function $G({\bf r}_1,{\bf r}_2)\equiv
\langle {\bf T}_\tau \psi ({\bf r}_1,\tau _1)\psi ^{\dagger }({\bf r}_2,\tau
_2)\rangle $ (where ${\bf T}_\tau $ means a thermal ordering in $\tau $) as
the ratio of two functional integrals,

\begin{equation}
\label{tcf}G({\bf r}_1,{\bf r}_2)=Z^{-1}\int e^S\psi ({\bf r}_1,\tau _1)\bar 
\psi ({\bf r}_2,\tau _2)D\psi D\bar \psi , 
\end{equation}
in which $Z$ is a partition function $Z=\int e^SD\psi D\bar \psi $. The
chosen c-number action $S$ is

\begin{equation}
\label{act}S=\int_0^\beta d\tau \int d^dr\left\{ \bar \psi ({\bf r},\tau
)K\psi ({\bf r},\tau )-\frac g2\bar \psi ({\bf r},\tau )\bar \psi ({\bf r}%
,\tau )\psi ({\bf r},\tau )\psi ({\bf r},\tau )\right\} . 
\end{equation}
The boundary conditions are vanishing at infinity for ${\bf r}$ and
periodic, period $\beta $, for $\tau $. The action $S$ yields the quantum
many-body problem at $T>0$ for a gas with repulsive pairwise $\delta $%
-function interactions of strength $g$ in ${\bf R}^d$. The differential
operator $K=\partial _\tau -H$, and $H=-\frac{\hbar ^2}{2m}\nabla ^2+V({\bf r%
})-\mu $; $\mu $ is the chemical potential and $V({\bf r})=\frac m2\Omega ^2%
{\bf r}^2$ is the paraboloidal trap potential, taken with spherical symmetry
for simplicity.

As explained we also simplify by setting $\psi ({\bf r},\tau )=\psi _o({\bf r%
})+\psi _1({\bf r},\tau )$ for c-number fields $\psi ,\psi _o,\psi _1$ and
likewise for $\bar \psi :$ $\psi _o({\bf r})$ will not depend on $\tau $. In
this Letter we shall only consider terms in $S$ upto quadratic (bilinear) in 
$\psi _1$,$\bar \psi _1$ and can therefore explicitly integrate out the
thermal fluctuations. This way we arrive at $S_{eff}[\psi _o,\bar \psi
_o]=\ln \int e^SD\psi _1D\bar \psi _1,$ and

\begin{eqnarray}
&&\ S_{eff}[\psi _o,\bar \psi _o]+\beta F_{nc}(\mu ) \nonumber \\
&=&\ \beta \int d^dr\{\bar \psi _o({\bf r})[\frac{\hbar ^2}{2m}\nabla ^2+\mu
-\tilde V({\bf r})]\psi _o({\bf r})-\frac g2\bar \psi _o({\bf r})\bar \psi
_o({\bf r})\psi _o({\bf r})\psi _o({\bf r})\}.
\end{eqnarray}
In $S_{eff},$ $\tilde V({\bf r})=V({\bf r})+2g\rho _{nc}({\bf r})$ is a
renormalised trap potential, while $\rho _{nc}({\bf r})$ is the density
profile of the thermal particles in the ideal gas approximation - as is
consistent with terms only quadratic in $\psi _1,\bar \psi _1$ retained.
More precisely, at this level of approximation (Hartree-Fock-Bogoliubov, HFB)
this density profile derives from the fundamental solution of the $d+1$%
-dimensional operator $K$: $K{\cal G}({\bf r},\tau ;{\bf r}^{\prime },\tau
^{\prime })=-\delta ({\bf r}-{\bf r}^{\prime })\delta (\tau -\tau ^{\prime })
$, and the thermal Green's function ${\cal G}$ can be expressed in the form

\begin{equation}
\label{igf}{\cal G}({\bf r},\tau ;{\bf r}^{\prime },\tau ^{\prime })=\sum_{%
{\bf n}}{\frac{u_{{\bf n}}({\bf r})u_{{\bf n}}({\bf r}^{\prime })}{e^{\beta 
{\rm E}_{{\bf n}}}-1}}e^{{\rm E}_{{\bf n}}(\tau -\tau ^{\prime })}, \tau >\tau ^{\prime }.
\end{equation}
For $d=3$ the vectors ${\bf n}=(n_1,n_2,n_3)$, and the $u_{{\bf n}}({\bf r})$
and ${\rm {E}_{{\bf n}}}$ are the eigenfunctions and eigenenergies,
respectively, of the $d=3$ harmonic oscillator Hamiltonian $H$; and
similarly for $d=2,d=1$. Then $\rho _{nc}({\bf r})={\cal G}^{\prime }({\bf r}%
,\tau ;{\bf r},\tau ),$ where prime means ${\bf n}=(0,0,0)$ is omitted. The
free energy of the thermal particles $F_{nc}(\mu )$ is simply $F_{nc}(\mu
)=\beta ^{-1}\sum_{{\bf n}}^{^{\prime }}\ln (1-e^{-\beta {\rm {E}_{{\bf n}}}%
})$.

At this point we are already able to calculate the critical temperatures $T_c$%
. The leading {\it i.e. }zeroth order term is found by replacing $\rho _{nc}(%
{\bf r})$ by a constant $\rho _{nc}(0)$: this defines the renormalised
chemical potential $\Lambda =\mu -2g\rho _{nc}(0)$, and $\Lambda =0$
determines $T_c$. We find this way all of the zeroth order expressions in $%
d=3,2,1$ as given e.g. in \cite{7}. At first order in $g$ we can add the
appropriate terms arising in $S_{eff}$. Beyond this we also need to include
fluctuations $\psi _1$ to an order higher than quadratic.

The free energy $F$ of the trapped Bose gas is calculated from $-\beta F(\mu
)=\ln \int e^{S_{eff}}D\psi _0D\bar \psi _0$. By steepest descents for large 
$\beta $ (low $T$) we find that 
\begin{equation}
\label{f}F(\mu )=F_{nc}(\mu )-\frac g2\int d^dr\mid \Phi ({\bf r})\mid ^4.
\end{equation}
In Eq.(\ref{f}) the fields $\Phi ,\bar \Phi $ are the quasi-classical fields
satisfying the extremum condition $\delta (S_{eff}[\Phi ,\bar \Phi ])=0$.
This condition is equivalent to the stationary GP equations

\begin{equation}
\label{gp}\frac{\hbar ^2}{2m}\nabla ^2\Phi ({\bf r})+(\mu -\tilde V({\bf r}%
))\Phi ({\bf r})-g\mid \Phi ({\bf r})\mid ^2\Phi ({\bf r})=0, 
\end{equation}
and the similar equation for $\bar \Phi $. At this quasi-classical
approximation we already find through the presence of $\tilde V({\bf r})$
the HFB corrections to the GP
equation introduced earlier \cite{7}.

We turn next to the calculation of the finite-temperature correlation
function Eq.~(\ref{tcf}). By integrating out the thermal fluctuations
included upto terms quadratic in $\psi _1$,$\bar \psi _1$ we find that

\begin{equation}
\label{cffi}G({\bf r}_1,{\bf r}_2)\simeq {\frac{\int e^{S_{eff}}\psi _o(%
{\bf r}_1)\bar \psi _o({\bf r}_2)D\psi _oD\bar \psi _o}{\int e^{S_{eff}}D\psi
_oD\bar\psi _o}\equiv C(}{\bf r}_1,{\bf r}_2).
\end{equation}
At low enough temperatures these remaining functional integrals can be
evaluated by steepest descents where again we work consistently at the HFB
level. At this level the correlation functions can be expressed in the form

\begin{equation}
\label{cffi2}C({\bf r}_1,{\bf r}_2)\simeq e^{-S_{eff}[\Phi _0,\bar \Phi
_0]+S_{eff}[\Phi _1,\bar \Phi _1]+\ln \Phi _1({\bf r}_1)\bar \Phi _1({\bf r}%
_2)},
\end{equation}
and the fields $\Phi _0,\bar \Phi _0$ satisfy the two stationary
GP equations exemplified by Eq.~(\ref{gp}). Evidently the fields $\Phi
_1,\bar \Phi _1$ are determined by $\delta (S_{eff}[\Phi _1,\bar \Phi
_1]+\ln \Phi _1({\bf r}_1)\bar \Phi _1({\bf r}_2))=0$, and this variational
equation also leads to a pair of GP type of equations with
however additional sources from the $\ln (\Phi _1\bar \Phi _1)$. This pair
of equations is

\begin{eqnarray}
-\frac{\hbar ^2}{2m}\nabla ^2\Phi _1({\bf r})-(\mu -\tilde V({\bf r}))\Phi
_1({\bf r})+g\Phi _1^2({\bf r})\bar \Phi _1({\bf r}) &=&{{\frac {\delta ({\bf r}-{\bf r}_2)}{\beta \bar 
\Phi _1({\bf r}_2)}}},  \label{gps} \\
-\frac{\hbar ^2}{2m}\nabla ^2\bar \Phi _1({\bf r})-(\mu -\tilde V({\bf r}))%
\bar \Phi _1({\bf r})+g\bar \Phi _1^2({\bf r})\Phi _1({\bf r}) &=&{{\frac {\delta ({\bf r}-{\bf r}_1) }{\beta \Phi _1({\bf r}_1)}}}.  \nonumber
\end{eqnarray}
In this Letter, for simplicity, we shall solve both these equations Eqs.~(\ref{gps})
and the Eq.~(\ref{gp}) only at the
Thomas-Fermi (TF) approximation: this is expected to be valid at low enough
temperatures \cite{7,8,9}. The stationary GP equation
Eq.~(\ref{gp}) is thus reduced to a simple algebraic equation and from it we then
easily find the expected inverted paraboloidal density profile which is \cite{7,hau}

\begin{equation}
\label{tfs}\rho _0({\bf r})\equiv \Phi _0({\bf r})\bar \Phi _0({\bf r})=%
\frac 1g(\mu -\tilde V({\bf r}))\Theta (\mu -\tilde V({\bf r})),
\end{equation}
in which $\Theta $ is the Heaviside step function. Evidently $\rho _0({\bf r}%
)$ can be interpreted as the condensate density expressed in terms of order
parameters $\Phi _0,\bar \Phi _0$. The radius of the condensate $R_c$ can
now be determined from the condition $\mu -\tilde V(R_c)=0$. 

The solution of Eqs.~(\ref{gps}) is more complicated. Notice first that the
fields $\Phi _1,\bar \Phi _1$, appearing in these equations will also be complex
valued fields in general. But they are two independent fields with
independent variations. We can therefore seek first of all solutions in the
form
$\Phi _1({\bf r})=\sqrt{\rho ({\bf r})}e^{\phi ({\bf r})},\,\,\bar \Phi _1(%
{\bf r})=\sqrt{\rho ({\bf r})}e^{-\phi ({\bf r})},$
where $\phi ({\bf r})$ can be complex valued, but will be found below to
describe the real contribution to the correlation functions of the complex
phases of the wave functions. We can furthermore assume that, away from the
boundaries, $\rho ({\bf r})$ is a slowly varying function of position ${\bf r%
}$ so that $\nabla ^2\sqrt{\rho }$ and $\nabla \sqrt{\rho }$ are both small
and can be neglected. This will not be true of $\rho (\nabla \phi )^2$ or $%
\rho \nabla ^2\phi $, so that Eqs.~(\ref{gps}) can be taken as

\begin{equation}
\label{gps1}g\rho ({\bf r})-(\mu -\tilde V({\bf r}))-\frac{\hbar ^2}{2m}%
(\nabla \phi ({\bf r}))^2=0
\end{equation}
\begin{equation}
\label{gps2}\frac{\hbar ^2}{2m}\nabla ^2\phi ({\bf r})={{\frac 1{2\beta \rho
({\bf r}_1)}}}\delta ({\bf r}-{\bf r}_1)-{{\frac 1{2\beta \rho ({\bf r}_2)}}}%
\delta ({\bf r}-{\bf r}_2).
\end{equation}
The first of these equations Eq.~(\ref{gps1}) has the solution $\rho ({\bf r}%
)=\rho _0({\bf r})+\frac{\hbar ^2}{2mg}(\nabla \phi )^2$. Within the
TF approximation, $\rho ({\bf r})$ in Eq.~(\ref{gps2}) is then $%
\rho _0({\bf r})$. We then express the solution of this equation in terms of
a function $f({\bf r},{\bf r}^{\prime })$:
$\phi ({\bf r};{\bf r}_1,{\bf r}_2)=f({\bf r},{\bf r}_1)-f({\bf r},{\bf r}%
_2).$
The functional form of $f({\bf r},{\bf r}^{\prime })$ depends on the
dimensionality of the system. We find that

\begin{equation}
\label{f1}f({\bf r},{\bf r}^{\prime })=-\frac a{2\pi \beta \rho _0({\bf r}%
^{\prime })}\ \frac 1R\,\,\,\,(d=3)
\end{equation}
\begin{equation}
\label{f2}f({\bf r},{\bf r}^{\prime })=\frac a{\pi \beta \rho _0({\bf r}%
^{\prime })}\ln R\ \ (d=2)
\end{equation}
\begin{equation}
\label{f3}f({\bf r},{\bf r}^{\prime })=\frac a{\beta \rho _0({\bf r}^{\prime
})}R\ \ (d=1)
\end{equation}
with $a\equiv \frac m{2\hbar ^2}$ and $R\equiv \mid {\bf r}-{\bf r}^{\prime
}\mid $. It is already clear that the correlation functions can no longer
depend on $R=\mid {\bf r}-{\bf r}^{\prime }\mid $ alone: they depend also on
both of ${\bf r}_1$ and ${\bf r}_2$ separately, consistent with the
breakdown of translational invariance induced by the trap.

We consider first the correlation function in $d=3$. In this case the points 
${\bf r=r}_1$ and ${\bf r=r}_2$ in $\phi ({\bf r};{\bf r}_1,{\bf r}_2)$ are
singular and introduce a divergence problem \cite{foot}. This difficulty can
be avoided first of all by considering a first-order 'coherence function'  $%
G^{(1)}({\bf r}_1,{\bf r}_2)$ which we define here (and compare e.g. \cite{Glauber,loud}) 
as

\begin{equation}
\label{cohf}G^{(1)}({\bf r}_1,{\bf r}_2)=\frac{G({\bf r}_1,{\bf r}_2)}{%
\langle \psi ({\bf r}_1,\tau _1)\rangle \langle \psi ^{\dagger }({\bf r}%
_2,\tau _2)\rangle }\simeq \frac{C({\bf r}_1,{\bf r}_2)}{\langle \psi _o(%
{\bf r}_1)\rangle \langle \bar \psi _o({\bf r}_2)\rangle },
\end{equation}
where $C$ was defined at Eq.~(\ref{cffi}). For this way $G^{(1)}$ is both finite and
well defined because we find identically the same singularities appear 
\cite{pop1} in the direct
calculation of the order parameters $\langle \psi ({\bf r}_1,\tau
_1)\rangle ,\langle \psi ^{\dagger }({\bf r}_2,\tau _2)\rangle $. Notice
that in Eq.~(\ref{cohf}) we have already replaced $\langle \psi \rangle $,
the order parameter  of the trapped Bose gas by $\langle \psi _o\rangle $
since the average over thermal fluctuations vanishes: $\langle \psi _1\rangle =0$%
. Notice too that the solution of Eq.~(\ref{gps2}) is only determined within a
harmonic function, and that the only harmonic function which satisfies the
b.c.s. and is non-singular is a (complex) number. This constant is
identically cancelled in $G^{(1)}$ but introduces an undetermined phase, as
expected  \cite{dun,m,leg}, to the order parameters. For $T<T_c$, when the order
parameter is nonzero, we find that

\begin{equation}
\label{g3}G^{(1)}({\bf r}_1,{\bf r}_2)\simeq e^{-\frac 12(f({\bf r}_1,{\bf r}%
_2)+f({\bf r}_2,{\bf r}_1))}=e^{\frac a{4\pi \beta \bar \rho _0}\frac 1R},
\end{equation}
where $\bar \rho _0^{-1}\equiv \rho _0^{-1}({\bf r}_1)+\rho _0^{-1}({\bf r}%
_2)$. Evidently $G^{(1)}({\bf r}_1,{\bf r}_2)\rightarrow 1$ for large $R
$, thus indicating long-range order and long-range coherence, and there are
thus these features of a coherent state in this limited sense \cite{note}. 
The coherence length is
given by $\frac a{4\pi \beta \bar \rho _0}$ and depends on both ${\bf r}_1$
and ${\bf r}_2$ separately. Notice that we have assumed ${\bf r}_1$ and $%
{\bf r}_2$ are not close to the boundaries of the condensate so that  always 
$\rho _0({\bf r})>0$ in $\bar \rho _0^{-1}$.

For $d=2$ the singularity in $f({\bf r},{\bf r}^{\prime })$ is logarithmic
and the divergence is renormalizable. For $d=1$ the function $f$ is
nonsingular. Thus we can directly evaluate the correlation functions, and
find that

\begin{equation}
\label{g2}G({\bf r}_1,{\bf r}_2)\simeq \sqrt{\rho _0({\bf r}_1)\rho _0({\bf r%
}_2)}e^{-\frac a{2\pi \beta \bar \rho _0}\ln R}\ \ (d=2),
\end{equation}
\begin{equation}
\label{g1}G({\bf r}_1,{\bf r}_2)\simeq \sqrt{\rho _0({\bf r}_1)\rho _0({\bf r%
}_2)}e^{-\frac a{2\beta \bar \rho _0}R}\ \ \ (d=1).
\end{equation}
It is obvious that these correlation functions both vanish for large $R$ and
that there is no long-range order in $d=2$ or in $d=1$ nor will there be any
coherence. In the case $d=2$ the condensate is marginally stable in that
correlations decay algebraically, namely by a power law. The exponent of
this power-law is proportional to $T$ so that at very low temperatures,
correlations may thus prevail over almost macroscopic distances. In real
magnetic traps for $d=3$ we still expect the condensate to be stable
even for extremely anisotropic trap potentials, as in e.g. the experiments 
\cite{4} on atomic hydrogen. The three correlation functions Eqs.~(\ref{g3}-%
\ref{g1}) coincide with those obtained under translational invariance
without the trap to the extent that for $\Omega \rightarrow 0,V({\bf %
r)\rightarrow }0$, and we can expect $2g\rho _{nc}({\bf r})\rightarrow $Const%
$=2g\rho _{nc}$.

Thus in summary we have demonstrated that the functional integration
techniques developed for translationally invariant Bose systems can be
extended to Bose gases in a confining trap potential, and form also in this
case of a trap a convenient framework in which to consider the thermal
properties of the condensate. We have shown in particular that true
long-range order only arises in $d=3$ and that the condensate then appears 
asymptotically to be
in the coherent state essential for atom lasers. We expect to define
multi-point correlation functions similarly to Eq.~(\ref{cohf}) and find
these $\sim 1$ for large separations of the points.

One of us, N.M.B., would like to thank the
Department of Physics, University of Jyv\"askyl\"a
for support.
This work was partially supported by the Russian Foundation for Fundamental
Research. Grant 98-01-00313.


\begin{thebibliography}{**}
\bibitem{foot}  This divergence is well known in electrostatics and in
field theory as discussed e.g. in \cite{pop1}.
\end{thebibliography}
\end{document}